\documentclass[%
 aip,
 amsmath,amssymb,
 reprint,%
apl]{revtex4-1}

\usepackage{graphicx}
\usepackage{dcolumn}
\usepackage{bm}

\usepackage[utf8]{inputenc}
\usepackage[T1]{fontenc}
\usepackage{mathptmx}
\usepackage{etoolbox}

\usepackage{xcolor}
\DeclareRobustCommand{\mh}[1]{{\color{black}#1}}

\makeatletter
\def\@email#1#2{%
 \endgroup
 \patchcmd{\titleblock@produce}
  {\frontmatter@RRAPformat}
  {\frontmatter@RRAPformat{\produce@RRAP{*#1\href{mailto:#2}{#2}}}\frontmatter@RRAPformat}
  {}{}
}%
\makeatother
\begin{document}

\title{Real-time dynamics of VCMA-assisted switching of magnetic tunnel junctions}

	\author{Marco Hoffmann}
	\email[]{marco.hoffmann@mat.ethz.ch}
	\affiliation{Department of Materials, ETH Zurich, 8093 Zurich, Switzerland}
	
	\author{Shaohai Chen}
	\affiliation{Institute of Materials Research and Engineering (IMRE), Agency for Science, Technology and  Research (A*STAR), Singapore}
        \author{Gunasheel Kauwtilyaa Krishnaswamy}
        \affiliation{Department of Physics, National University of Singapore, Singapore}
        
        \author{Hang Khume Tan}
        \affiliation{Institute of Materials Research and Engineering (IMRE), Agency for Science, Technology and Research (A*STAR), Singapore}
        
        \author{Sherry L.~K.~Yap}
        \affiliation{Institute of Materials Research and Engineering (IMRE), Agency for Science, Technology and Research (A*STAR), Singapore}

        \author{James Lourembam}
        \affiliation{Institute of Materials Research and Engineering (IMRE), Agency for Science, Technology and Research (A*STAR), Singapore}
        
        \author{Anjan Soumyanarayanan}
	\email[]{anjan@nus.edu.sg}
	\affiliation{Institute of Materials Research and Engineering (IMRE), Agency for Science, Technology and  Research (A*STAR), Singapore}
    \affiliation{Department of Physics, National University of Singapore, Singapore}
    
	\author{Pietro Gambardella}
	\email[]{pietro.gambardella@mat.ethz.ch}
	\affiliation{Department of Materials, ETH Zurich, 8093 Zurich, Switzerland}
	
	\date{\today}
	
	\begin{abstract}

    Voltage control of magnetic anisotropy (VCMA) induced by charge accumulation is typically considered as an ultrafast process, enabling energy-efficient and high-speed magnetization switching in spintronic devices. In this work, we investigate the real-time dynamics of VCMA-assisted switching of magnetic tunnel junctions via relaxation in a magnetic field. We show that device-dependent charging effects and magnetic granularity in the free layer limit the switching speed at applied voltages close to the critical switching threshold. Increasing the voltage or the applied magnetic field reduces the incubation delay and total switching time to below a few ns. Micromagnetic simulations incorporating the finite charging times of the tunnel junction and the granularity of the magnetic film reproduce the experimental results, providing critical insights into optimizing VCMA-driven magnetization control for memory and logic applications. 
    	
    \end{abstract}
	
    \maketitle

Magnetization control via electric fields enables reversible manipulation of energy barriers at low power \cite{Weisheit2007,Maruyama2009,Endo2010,Shiota2012,Wang2012, Miwa2019,Shao2023a}. This voltage control of magnetic anisotropy (VCMA) in magnetic tunnel junctions (MTJs) \cite{Kanai2012, Nozaki2016} is critical to improve energy efficiency in magnetic random access memories \cite{Amiri2015, Andrawis2018, Nozaki2019, Krizakova2021, Krizakova2022, Shao2023a}. It is also relevant in MTJ-based random number generators and probabilistic bits \cite{Lee2017, Liu2022, Raimondo2024, Shao2023b}. Furthermore, magnetic domain walls \cite{Chiba2012, Bernand-Mantel2013}, skyrmions \cite{ Bhattacharya2020, Zhou2021, Chen2024, Urrestarazu2024} and spin waves \cite{Verba2017, Rana2018} can be created and controlled via VCMA.

In nm-thick metallic films such as the free layers of MTJs, VCMA is primarily induced through electrostatic charging of the metal/insulator interface, which changes the electronic occupancy of the $d$-orbitals and, consequently, the interfacial magnetic anisotropy energy of the magnetic layer \cite{Weisheit2007,Daalderop1991}. This effect does not break the time-reversal symmetry required for magnetization switching, but causes a reduction of coercive field. In combination with external fields or spin torques, VCMA can thus be used to facilitate magnetization reversal in MTJs \cite{Wang2012, Amiri2013, Kanai2014, Grimaldi2020,Krizakova2021, Wu2021}. In a macrospin system, sudden lifting of the magnetic anisotropy causes magnetization precession about an in-plane magnetic field. Precisely tuning the duration of the voltage pulse relative to the strength of the in-plane field allows for stopping the precession at the desired magnetization orientation \cite{Kanai2012, Shiota2012, Grezes2016}. Alternatively, VCMA can trigger incoherent magnetic relaxation and switching in MTJs biased by magnetic fields and/or spin torques using either unipolar or bipolar voltage schemes \cite{Wang2012, Kanai2014}. 
These various types of VCMA-assisted magnetic relaxation offer a practical approach for device implementation, as they do not require coherent magnetization reversal and sub-ns accuracy in pulse duration and rise time, which are key to reduce the write error rate in precessional switching \cite{Amiri2013, Yamamoto2019}. However, the transient dynamics of VCMA-assisted relaxation remains unexplored.

Although the alteration of the perpendicular magnetic anisotropy due to electron accumulation/depletion in a magnetic layer occurs on picosecond timescales, VCMA-assisted relaxation relies on thermal excitations, which inevitably introduce stochastic delays in magnetic switching \cite{Yamamoto2018, Kanai2021}. Furthermore, in MTJs the applied voltage does not instantaneously drop across the tunnel barrier due to the pseudocapacitive nature of the junction \cite{Padhan2007, Sahadevan2012, Kaiju2015, Ogata2021} and surrounding passivation layers \cite{Safeer2024}. These capacitive effects become more pronounced in large resistance-area product (RA) devices, required for VCMA-assisted switching at low current amplitudes.
    
Here, we report time-resolved measurements of VCMA-assisted switching in a perpendicular magnetic field for MTJs with large RA. While previous works have focused on determining switching probability after application of voltage pulses of different durations, we show that the underlying magnetization dynamics may evolve over much longer timescales than the pulse length. Switching delays can occur due to thermal incubation times, magnetic granularity, and capacitive electrostatic charging. For low applied voltages close to the critical $50\%$ switching threshold, magnetic reversal is triggered at the end of the voltage pulse and completed in the applied magnetic field in zero voltage conditions. The overall switching speed increases exponentially with applied voltage and field. Increasing the voltage and/or field also significantly narrows the statistical distribution of incubation times. These observations are corroborated by micromagnetic simulations that take into account capacitive charging of the junction and spatial variations of the magnetic anisotropy in the free layer.


    \label{section_samples_technique}

    \begin{figure}
		\includegraphics[width=85mm]{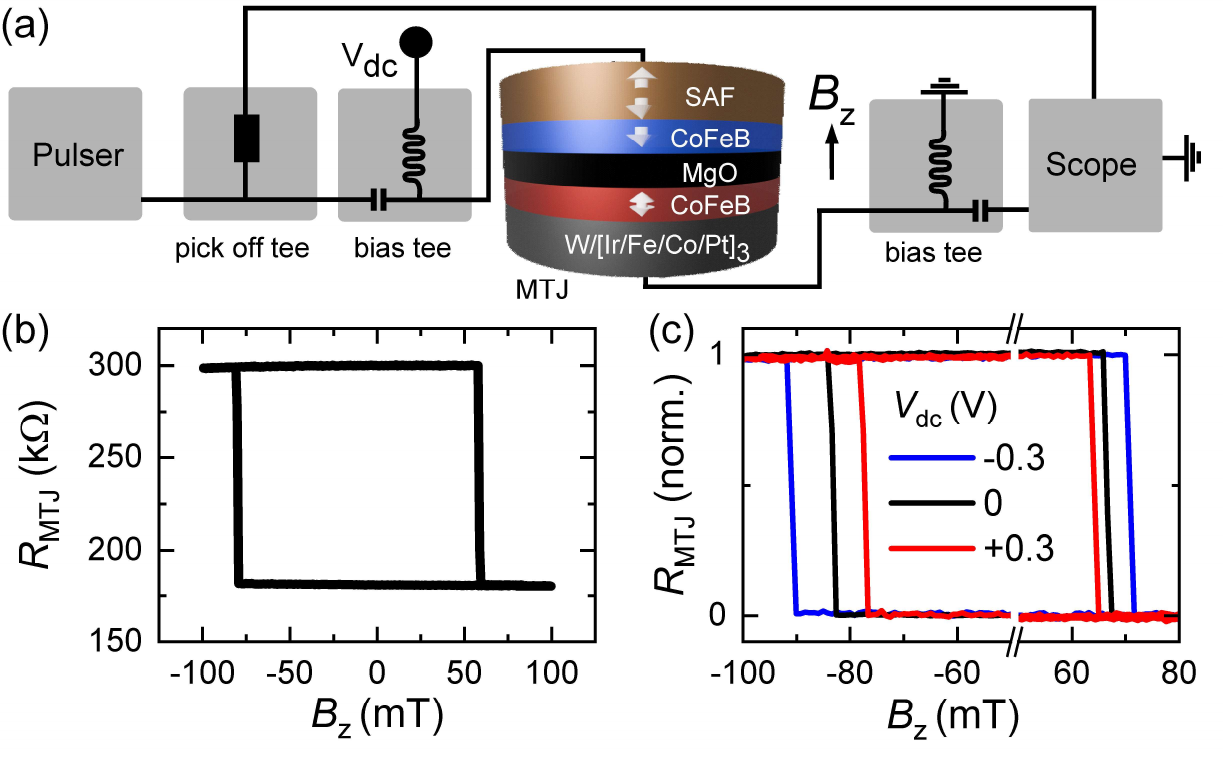}
		\caption{\label{fig1:Exp} (a) Schematic of the MTJ stack and experimental setup used for the dc- and time-resolved measurements. \mh{A pulse generator sends a voltage pulse that is probed by a pick-off tee before being fed to the top MTJ electrode. A bias-tee allows for dc-biasing of the junction.} (b) Hysteresis loop of an exemplary MTJ demonstrating typical TMR, coercivity and resistance of the investigated devices. (c) Impact of applied voltage $V_\text{dc}$  on the coercivity of the junction due to the VCMA effect.
        }
    \end{figure}
    
    Our MTJ devices comprise \mh{ $\mathrm{Co}_{20}\mathrm{Fe}_{60}\mathrm{B}_{20}$} magnetic free and reference layers with perpendicular magnetic anisotropy that are 0.9 and 1.3 nm thick, respectively. The MgO tunneling barrier is 2.2 nm thick and has an RA of about $\sim$24~k$\Omega \mu\text{m}^2$. The nominal diameter of the MTJ pillar is 320\,nm. The reference layer is top-pinned by a synthetic antiferromagnet (SAF), whereas the free layer is exchange-coupled to a $\text{[Ir/Fe/Co/Pt]}_3$ underlayer. This underlayer is used to tune the Dzyaloshinskii–Moriya interaction (DMI) of the bottom stack of the MTJ~\cite{Chen2024}, relevant to modulating the magnetic configuration of free layers with low anisotropy, a subject of future work. The MTJ devices selected for this study present uniform out-of-plane magnetization due to strong perpendicular magnetic anisotropy, and the DMI is not relevant for VCMA-assisted switching in a magnetic field as investigated here. More details on the stack and sample fabrication, which used a standard 200~mm back-end-of-line process are provided in Ref.~\cite{Chen2024}.

    The experimental setup, shown in Figure~\ref{fig1:Exp}(a), has a bandwidth of about 2.5~GHz. A pulse generator sends a square voltage pulse to the top electrode of the MTJ and the transmitted pulse is recorded by an oscilloscope. A pick-off tee allows for recording the input pulse while bias tees placed before and after the MTJ enable dc resistance measurements. We measure both the post-pulse switching probability and the real-time voltage during the application of each pulse. 
    The magnetic field $B_{\rm z}$ is applied perpendicular to the MTJ stack. 	
    Figure~\ref{fig1:Exp}(b) shows a representative hysteresis loop with two well-defined electrical states for parallel and antiparallel orientation of the free and reference layers, with a tunneling magneto-resistance (TMR) of about 70\%. The coercive field $B_\text{c}$ can be reduced or enhanced via VCMA using positive or negative applied voltages $V_\text{dc}$, respectively, as illustrated in Fig.~\ref{fig1:Exp}(c).

    \label{section_results_pp}
    	
    \begin{figure} [b]
		\includegraphics[width=85mm]{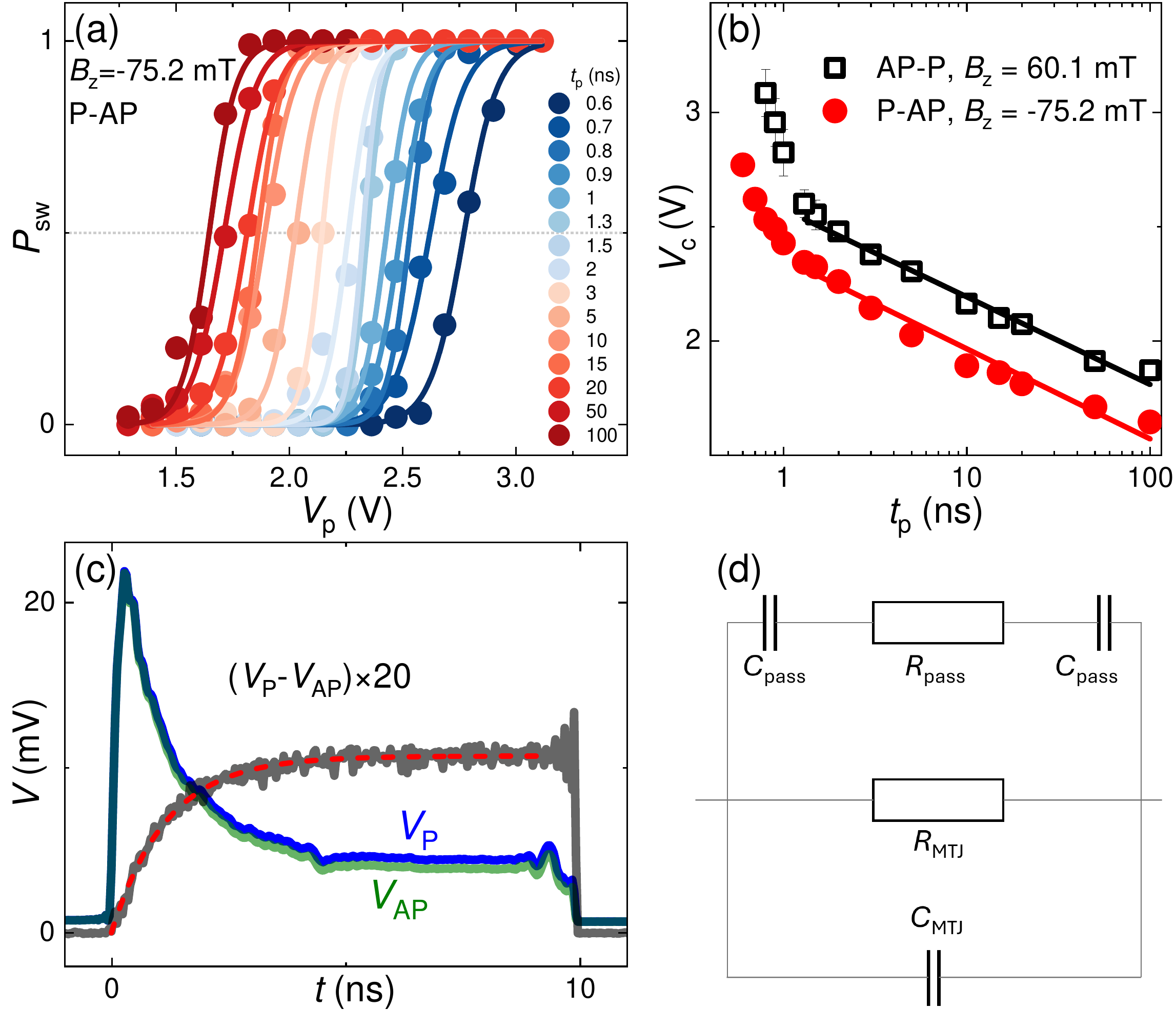}
		\caption{\label{fig2:pp} (a) Switching probability as a function of $V_\text{p}$ for several $t_\text{p}$ for the P-AP transition at $B_\text{z}= -75.2$\,mT. Each data point is the result of 100 switching attempts. The magnetization was reset to the P-state before every attempt. (b) Critical switching voltage as a function of $t_\text{p}$ for both switching directions. The lines show logarithmic fits for $t_\text{p}>$1.5\,ns. (c) Raw voltage traces recorded in the oscilloscope for pulses that do not induce switching in the P (blue) and AP (green) state stabilized by $B_\text{z}= \pm 100$\,mT. The black line is the difference of these two traces magnified by 20 and the red dashed line is an exponential fit. (d) Equivalent circuit of the MTJ showing the junction resistance, capacitance and passivation layer capacitances.}
	\end{figure}
    
    To perform VCMA-induced switching, we first saturate the magnetization of the free layer by applying a magnetic field $B_\text{z} = \pm100$\,mT. Subsequently, we set $B_\text{z}$ to a value close to $\mp B_\text{c}$ and check the TMR to ensure that the initialized uniform state is preserved. Then, a positive voltage pulse of amplitude $V_\text{p}$ and duration $t_\text{p}$ is applied to the top electrode to reduce the magnetic anisotropy and trigger magnetization reversal. \mh{While switching in an external field is impractical for applications, this procedure allows us to study the VCMA-induced magnetization dynamics, which is relevant also for devices switchable by all-electrical means \cite{ Wang2012, Grimaldi2020, Krizakova2021, Wu2021, Chen2024}.} Due to the high junction resistance exceeding 175~k$\Omega$, the current density passing the junction is about $10^{8}\text{A/m}^2$. This is two orders of magnitude lower than the typical critical current density required for STT switching. \mh{Therefore, VCMA-induced effects dominate over STT and associated Joule heating in our devices.}

 \begin{figure*}[ht!]
		\centering
		\includegraphics[width=170mm]{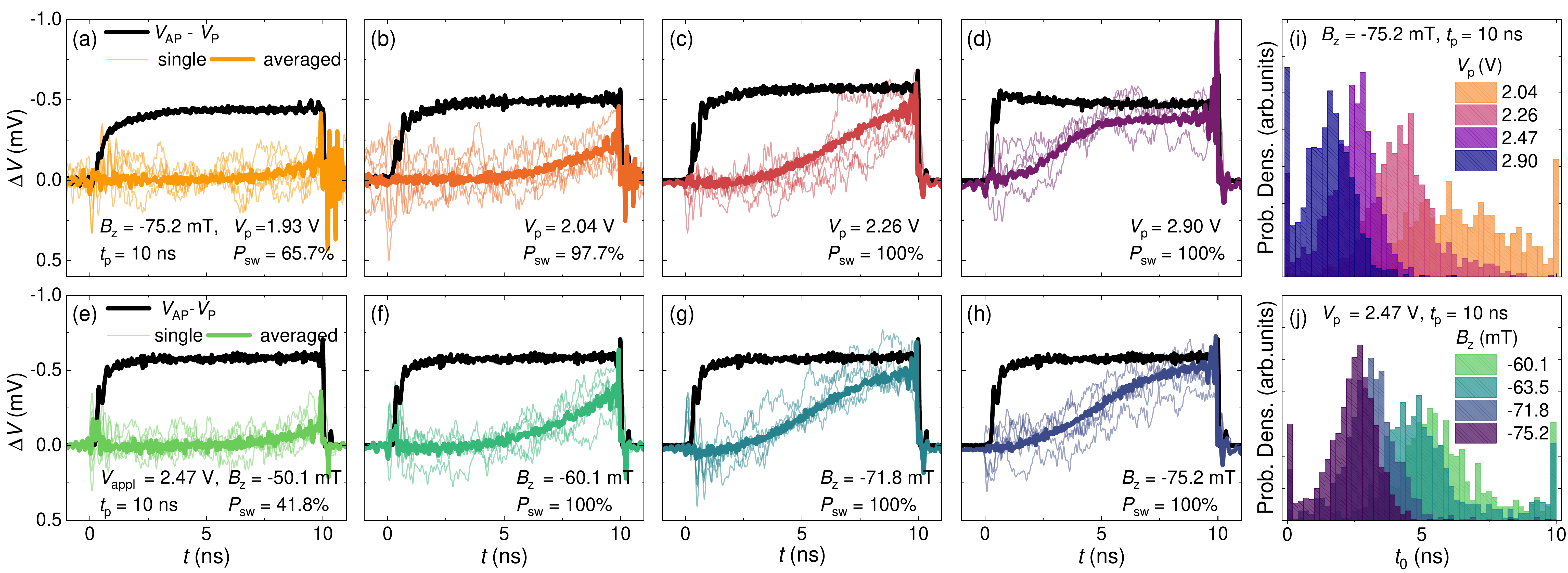} 
        \caption{\label{fig3:tra} Representative time-resolved traces for successful P-AP switching events. Single-shot (colored, thin) and averaged (colored, bold) traces are shown for varying values of $V_\text{p}$ (a-d) and $B_\text{z}$ (e-h).  
        The bold black trace is the difference of the voltage traces measured for fixed AP and P states, representing the full switching amplitude [see Fig.~\ref{fig2:pp}(c)]. Increasing switching probabilities are indicated from left to right. (i) Statistical distribution of the incubation time for varying $V_\text{p}$ (i) and $B_\text{z}$ (j) obtained by fitting sigmoid functions to the single-shot switching traces.  
        }
	\end{figure*}

    Figure~\ref{fig2:pp}(a) shows the switching probability $P_\text{sw}$ from the parallel to antiparallel (P-AP) state, computed from 100 switching attempts, as a function of $V_\text{p}$ for several $t_\text{p}$ at $B_\text{z}=-75.2$~mT. The solid lines are sigmoidal fits used to extract the critical switching voltage $V_\text{c}$ defined as the $V_\text{p}$ yielding $50\%$ successful reversals. $t_\text{p}$ and $B_\text{z}$ values yielding higher $P_\text{sw}$ are referred to as overcritical. $V_\text{c}$ is plotted versus $t_\text{p}$ for both P-AP and AP-P switching directions in Fig.~\ref{fig2:pp}(b). For $t_\text{p} > 1$~ns, $V_\text{c}$ decreases logarithmically with $t_\text{p}$ (solid lines), consistently with thermally-activated magnetization reversal \cite{Bedau2010, Alzate2012, Amiri2013, Garello2014}. 
    Negative voltage pulses do not cause switching, confirming that Joule heating does not play a significant role in our experiments. The lower $V_\text{c}$ of P-AP transitions relative to AP-P transition is assigned to minor STT contributions and to the spatial distribution of the SAF stray field \cite{Devolder2016, Chen2024, Zhou2024}, which supports P-AP reversal by favoring the nucleation of AP domains close to the boundary of the free layer.
    
    Additionally, we observe that $V_\text{c}$ increases steeply for $t_\text{p} < 1$~ns. Such a behavior cannot be attributed to an intrinsic electronic VCMA effect, which is considerably faster \cite{Shiota2012}. We thus investigate the electrostatic charging of the junction by recording the time-resolved traces of the oscilloscope during the applied voltage pulses for fixed P- and AP-states, as shown in Fig.~\ref{fig2:pp}(c). In these measurements, a strong magnetic field $B_\text{z} = \pm 100$\,mT prevents the magnetization from switching. These traces are a combination of an exponentially decreasing current charging the capacitances in the circuit and an exponentially saturating tunneling current. This is expected from an MTJ device behaving like a leaky capacitor \cite{Sahadevan2012, Kaiju2015} and confirmed by SPICE simulations for the equivalent circuit shown in Fig.~\ref{fig2:pp}(d). Fitting the raw traces or the difference of these raw traces - non-vanishing due to the TMR - with an exponentially decaying or saturating function results in a characteristic charging time of 1.3~ns. Thus, the pronounced increase of $V_\text{c}$ at low $t_\text{p}$ is ascribed to the finite charging time of the MTJ, yielding a smaller anisotropy reduction than expected from the applied voltage.


    To investigate the transient evolution of the magnetization initiated by VCMA, we report in Fig.~\ref{fig3:tra} the time-resolved traces of \emph{successful} P-AP switching events for varying values of $V_\text{p}$ at constant $B_\text{z}$ (a-d) and for varying $B_\text{z}$ at constant $V_\text{p}$ (e-h). Each switching trace represents the difference $\Delta V$ of the voltage recorded during switching and the P-state reference trace. Representative single-shot switching traces are shown as thin lines and traces averaged over several hundred events are shown as bold lines. The differences of transmitted voltages in the P- and AP-states fixed by fields $|B_\text{z}|> 100$~mT, shown as bold black lines, represent the maximum possible amplitude of the switching signal as determined by the TMR of the junction [see also Fig.~\ref{fig2:pp}(c)]. Note that $\Delta V$ vanishes after pulse termination since the TMR signal cannot be detected in the absence of current. For each parameter set, $P_\text{sw}$ is indicated at the bottom of the corresponding panel in Fig.~\ref{fig3:tra}. These data show that VCMA-assisted switching in a magnetic field is characterized by incubation times as long as 10~ns. Moreover, in close to critical conditions, even if the switching probability determined by post-pulse measurements is around $50\%$, only small deviations of the magnetization from the initial state are observed at the end of the voltage pulse, as shown in Fig.~\ref{fig3:tra}(a) and (e). Thus, in critical conditions, the switch is initiated but not completed within the pulse duration. Finalization of the reversal during the voltage pulse is only achieved by increasing $V_\text{p}$ in the overcritical regime, as shown in  Fig.~\ref{fig3:tra}(d), or upon increasing $B_\text{z}$ towards $B_\text{c}$, as shown in Fig.~\ref{fig3:tra}(h). This behavior contrasts with that observed for spin torque switching, for which finalization of the reversal is always reached during the applied pulse \cite{Devolder2016, Grimaldi2020, Krizakova2021}.

Fitting the single-shot traces with sigmoidal functions, we define the incubation time $t_0$ as the time at which 10\% of the reversal has been completed. Figure \ref{fig3:tra}(i) and (j) show that the mean and standard deviation of $t_0$ reduce upon increasing $V_\text{p}$ and $|B_\text{z}|$, respectively, as expected due to the corresponding reduction of the reversal energy barrier. However, single-shot data around critical conditions [Fig. \ref{fig3:tra}(a) and (e)] do not yield reliable fits due to electrical noise superposed to the small changes of the magnetization. Note that, for VCMA-assisted switching of MTJs with large RA product, the noise in the voltage traces is considerably higher relative to that for spin torque switching measurements \cite{Devolder2016, Grimaldi2020, Krizakova2020APL, Krizakova2021}. \mh{A better signal-to-noise ratio may be obtained by lowering the RA and/or increasing the TMR ratio.}

\begin{figure}[hbt!]
		\centering
		\includegraphics[width=90mm]{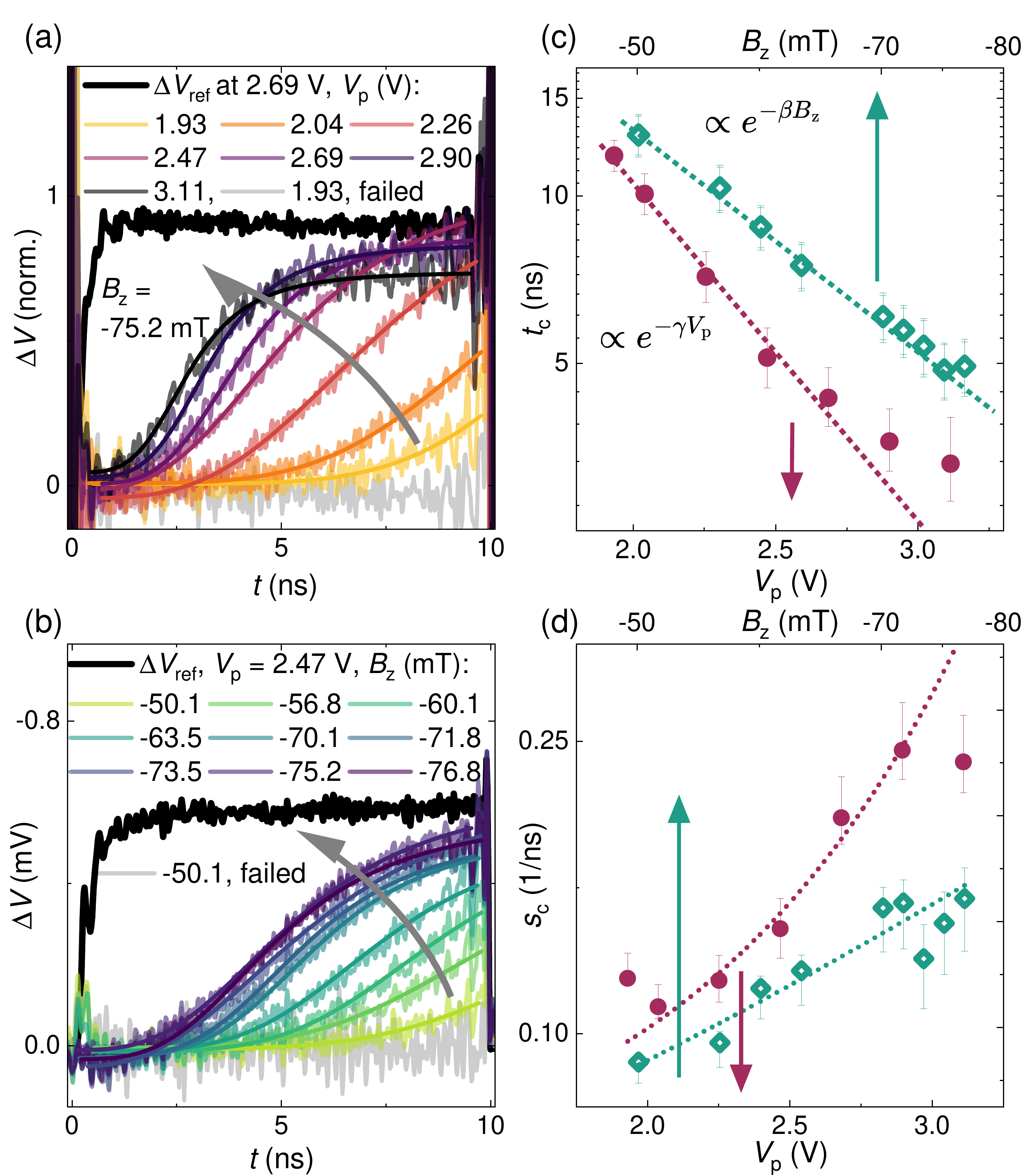} 
		\caption{\label{fig3:trb} Successful switching traces for varying $V_\text{p}$ (a) and $B_\text{z}$ (b)  averaged over 500 pulses. The solid lines are sigmoidal fits to the data. Gray traces indicate failed averaged traces. (c) Times at which $50\%$ of the switching is achieved according to the fits shown in (a,b). The data are plotted as a function of $V_\text{p}$ (circles) and $B_\text{z}$ (diamonds). 
        The dashed lines are exponential fits to the data. (d) Slope of the fitted sigmoidal functions at $50\%$ switching as a function of $V_\text{p}$ (circles) and $B_\text{z}$ (diamonds). The dashed lines are fits to the data (see text). The error bars in (c,d) indicate the slopes at $40\%$ and $60\%$ switching.
        }
    \end{figure}

To explore the incubation times and reversal speeds in the entire parameter range of our measurements, we analyze the averaged switching traces of the successful attempts shown in Fig.~\ref{fig3:trb}(a) and (b).
By fitting sigmoid functions to these traces, we extract the times $t_\text{c}$ at which the fits reach $50\%$ of the maximum signal and plot them as functions of $V_\text{p}$ and $B_\text{z}$ in Fig.~\ref{fig3:trb}(c). 
The dashed lines in Fig.~\ref{fig3:trb}(c) represent exponential fits to $e^{-\beta B_{\rm z}}$ (green line) and $e^{-\gamma V_{\rm p}}$ (red line), consistent with the reduction of the switching energy barrier induced by field and voltage, respectively \cite{Sala2022, Wu2022}. Here, the fit parameters $\beta$ and $\gamma$ characterize how the barrier decreases with $B_\text{z}$ and $V_\text{p}$. The Arrhenius-like reduction of the characteristic switching time $t_\text{c}$ is expected in both cases for switching assisted by random thermal fluctuations \cite{Sala2022}. The reduction in $t_\text{c}$ achieved by $V_\text{p}$ is steeper than the one by $B_\text{z}$.
This may be due to a nonlinear dependence of the switching energy barrier on $V_\text{p}$ or to small supportive STT and self-heating contributions to switching \cite{Krizakova2021}. However, the reduction of $t_\text{c}$ with $V_{\rm p}$ tails off below about 3~ns due to finite charging time of the MTJ. Comparing the changes in $B_\text{z}$ and $V_\text{p}$ that achieve a similar reduction in $t_\text{c}$, we estimate a field-to-voltage linearity factor $\delta B /\delta V \approx 36(5)$\,mT/V. This conversion ratio corresponds to an estimated VCMA coefficient $\xi = M_s t_\text{CoFeB} t_\text{MgO} \delta B /\delta V \approx 88(12)$~fJ/(Vm), where $M_s=1.24$~MA/m is the saturation magnetization of the free layer and $t_\text{CoFeB}$, $t_\text{MgO}$ are the thickness of the free layer and MgO barrier, respectively. This estimate is within the range of 25-100~fJ/(Vm) typically reported for CoFeB/MgO interfaces \cite{Nozaki2019} and comparable to the estimates based on dc measurements \cite{Chen2024}.

Similar to spin torque induced switching \cite{Devolder2016, Grimaldi2020,Baumgartner2017}, micromagnetic simulations indicate that the VCMA-assisted reversal process proceeds by domain nucleation and expansion. The slopes ($s_\text{c}$) of the averaged switching traces in Fig.~\ref{fig3:trb}(a) and (b) can thus serve as a proxy for the effective reversal rate. 
The values of $s_\text{c}$ evaluated at $50\%$ switching amplitude exhibit a characteristic increase as a function of $B_\text{z}$ and $V_\text{p}$ followed by saturation, as shown in Fig.~\ref{fig3:trb}(d). 
The dashed lines are fits proportional to $e^{-\beta^{\prime}B_\text{z}^{-1/4}}$(green line) and $ e^{-\gamma^{\prime}(1+\delta V_\text{p})^{9/8}}$ (red line), consistent with a model of VCMA-assisted creep motion of domain walls in a magnetic field, with effective magnetic anisotropy $K_\text{eff}$ and fitting constants $\beta^{\prime}$, $\gamma^{\prime}$, and $\delta$ \cite{Metaxas2007, LiuThesis2017, Liu2017jap}. Deviations of $s_\text{c}$ from the fits occur at high-field and high-voltage, likely due to the onset of charging effects.

To further investigate the dynamics of VCMA-assisted switching, we performed micromagnetic simulations using the MuMax3 code \cite{ Vansteenkiste2014}. First, an effective medium approximation \cite{Woo2016} was applied to model the entire MTJ stack as nine effective layers of equal thickness to extract the stray field of the SAF. Then, the magnetic free layer was modeled as a 320-nm-diameter disk, discretized into 192 $\times$ 192 $\times$ 1 cells and exposed to the SAF stray field.  The effective magnetic anisotropy was set to $K_\text{eff} = 0.54\,\text{MJ/m}^3$ and the CoFeB is modelled with an exchange stiffness  $A_\text{ex} = 24\,\text{pJ/m}$, saturation magnetization $M_\text{sat}=1.24\,\text{MA/m}$, and DMI amplitude $D=0.05\,\text{mJ/m}^2$ \cite{Chen2024}.   
Stochastic thermal effects were included via a fluctuating thermal field at 300\,K \cite{Vansteenkiste2014}. We simulated the time evolution of the free layer magnetization under voltage pulses that reduced the magnetic anisotropy, assuming a VCMA factor $\xi = 42\,\text{fJ/Vm}$ as derived from dc measurements for the investigated sample and typical for CoFeB-MgO interfaces \cite{Alzate2014}. The capacitive charging of the MTJ was also considered, introducing a characteristic rise time of 1.3~ns, as derived from the experiments [see Fig.~\ref{fig2:pp}(c)].   
    \begin{figure}
		\includegraphics[width=75mm]{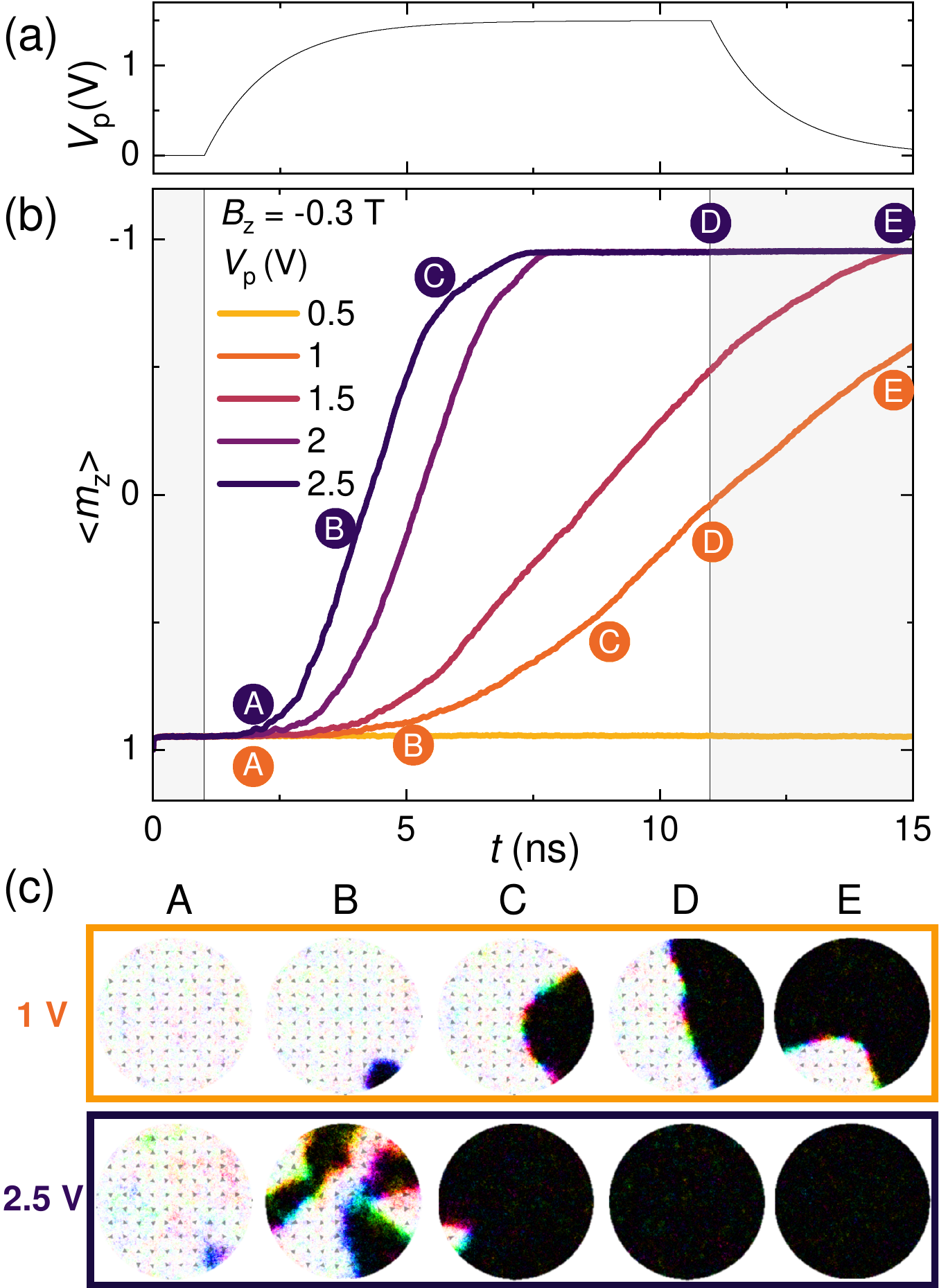}
		\caption{\label{fig4:sims} (a) Temporal profile of the voltage pulse used in the micromagnetic simuations. (b) Transient behavior of the $z$-component of the magnetization in response to voltage pulses of varying amplitude. The gray shaded areas indicate the time ranges inaccessible to real-time measurements before and after the pulse. (c) Spatial distribution of the magnetization during the reversal process at the times indicated by letters in (b) for slow (top row) and fast switching (bottom row).}
    \end{figure}
To model our real-time measurements, we conducted simulations varying $B_\text{z}$, $V_\text{p}$, temperature, and the variation of the magnetic parameters in the free layer over different crystal grains. We considered 50 different grains using Voronoi tesselation with an approximate average size of about 40~nm \mh{based on TEM images} and $K_\text{u}$, $M_\text{sat}$, $A_\text{ex}$, and $D$ normally distributed around their nominal value \mh{ with standard deviation $\sigma = 10$\%. Variations of $K_\text{u}$ and $M_\text{sat}$ have the largest impact on the $t_\text{0}$ and $t_\text{c}$.}
The exchange interaction between neighbouring grains was reduced by $10\%$ with respect to a uniform layer.
Introducing granularity was a key requirement for reproducing the experimental behavior across all simulation sets, in agreement with previous studies on similar samples \cite{Chen2024, Legrand2017, Bhattacharya2020, Jenkins2024}. 
In granular films, propagating domain walls can break and become pinned, slowing down the reversal process. In contrast, simulations of homogeneous magnetic layers showed full reversal within 2–3 ns, significantly faster than experimentally observed. 

Figure \ref{fig4:sims}(a) illustrates a 10-ns voltage pulse simulation, incorporating the finite charging time of the MTJ. The corresponding time evolution of the $z$-component of magnetization under pulses of varying amplitudes and a negative magnetic field is shown in Fig.~\ref{fig4:sims}(b). For undercritical $V_\text{p} = 0.5$~V (gray) no inverted domain is nucleated, and thus no switching occurs.
For the lowest $V_\text{p}$ achieving switching, magnetization reversal begins during the pulse but completes only after pulse termination, driven by $B_\text{z}$, in agreement with the experimental data in Fig.~\ref{fig3:tra}. As $V_\text{p}$ increases or $B_\text{z}$ approaches $B_\text{c}$, a larger fraction of the reversal occurs within the pulse duration, leading to complete magnetization switching within a few nanoseconds. Figure \ref{fig4:sims}(c) presents magnetization maps for two distinct cases: a slow reversal (upper row) at low $V_\text{p}$, and a fast reversal (lower row) at high $V_\text{p}$. Reversal typically nucleates at grains with lower $K_\text{u}$ and then expands. Larger voltage pulses, which induce a larger reduction in magnetic anisotropy during the pulse, facilitate multiple nucleation sites and thus accelerate the completion of the reversal. \mh{Simulations performed for different values of $\sigma$ and grain sizes yield qualitatively similar conclusions. However, increasing $\sigma$ between 5\% and 15\% leads to more than 50\% reduction of $t_\text{0}$ and $t_\text{c}$ as nucleation is favored in the grains with weaker magnetic anisotropy. Varying the grain size from 30 to 60~nm leads to 5\%-25\% reduction of $t_\text{0}$ and $t_\text{c}$ for varying $V_\text{p}$ with the largest changes at $V_\text{c}$.}


In summary, we performed real-time measurements of VCMA-assisted magnetization switching in a magnetic field within highly resistive MTJs. While the VCMA effect due to charge accumulation is inherently ultrafast, the junction charging time extends to about 1~ns, and the magnetization response unfolds over several nanoseconds, depending on the applied voltage and field.
We identify a switching incubation delay associated with stochastic thermal activation, with a lower bound set by the junction charging time. Under near-critical conditions, magnetization reversal is initiated during a 10-ns-long voltage pulse, and completes only after pulse termination via domain wall creep. Increasing the applied voltage proves more effective than increasing the magnetic field in reducing both the incubation time and overall switching time. \mh{Further improvements in the switching dynamics may be obtained by minimizing the stray capacitances due to the substrate and passivation layer surrounding the MTJ.}
Micromagnetic simulations reproduce the experimental switching traces, revealing that domain wall pinning within grains of varying magnetic anisotropy dominates the post-incubation reversal dynamics at low voltages. Our findings provide insights into VCMA-assisted switching in the absence of significant self-heating and spin torques. In practical applications, they are relevant for VCMA devices where precise control over the switching dynamics is required to prevent overlap between sequential switching operations.
\begin{acknowledgments}
This project was supported by the European Union’s Horizon 2020 Research and Innovation Programme under the Marie Skłodowska-Curie grant agreement N. 955671 and the Swiss National Science Foundation (Grant N. 200021-236524). This work was supported by Singapore’s Manufacturing, Trade and Connectivity (MTC) initiative (A*STAR Grant No. M23M6c0112), the SpOT-LITE programme (A*STAR Grant No. A18A6b0057), and the Ministry of Education Academic Research Fund (Tier-1 NUS Grant No. 25-0760-A0001).
\end{acknowledgments} 
\section*{Conflict of interest}
The authors have no conflicts to disclose.
\section*{Author contributions}
M.H.: Measurements, analysis, data curation, software, visualization, writing
S.C: project design, conceptualization, investigation
G.K.: project design, conceptualization, investigation
H.K.T. stack fabrication
S.L.L.Y.: device fabrication
J.L.: stack and process design, review and editing
A.S.: Funding acquisition, supervision, review and editing
P.G.: Funding acquisition, methodology, supervision, writing, review and editing

\section*{Data Availability Statement}
The data that support the findings of this study are made openly available in the ETH Research Collection using the DOI 10.3929/ethz-b-000742264.

\bibliography{bib_pmtj_v1}
 \newpage
 \clearpage
 \onecolumngrid

\end{document}